\documentstyle[psfig]{article}
\newcommand{\thetitle}{}
\newcommand{\thesubtitle}{}
\newcommand{\thedate}{}
\newcommand{\theauthor}{}
\newcommand{\theinstitute}{}

\renewcommand{\title}[1]{\renewcommand{\thetitle}{#1}}

\renewcommand{\date}[1]{\renewcommand{\thedate}{#1}}
\renewcommand{\author}[1]{\renewcommand{\theauthor}{#1}}
\newcommand{\institute}[1]{\renewcommand{\theinstitute}{#1}}

\newcommand{\thesaurus}[1]{}
\newcommand{\offprints}[1]{\renewcommand{\thefootnote}{}
\footnotetext{{\it Send offprints
requests to:} #1}\renewcommand{\thefootnote}{\arabic{footnote}}
}

\renewcommand{\maketitle}{
\thispagestyle{empty}
{\parindent0cm
\begin{center}
\vskip 2em
{\LARGE \thetitle \par {\Large \thesubtitle}}\par
\vskip 2em {\large \theauthor}\par
\vskip 1em {\theinstitute}\par
\vskip 2em {\thedate}
\end{center}
}}

\newcommand{\sun}{\mbox{$\odot$}}
\newcommand{\la}{\le}
\newcommand{\ga}{\ge}
\newcommand{\keywords}{{\bf Keywords:} }

\newcommand{\degr}{\mbox{$^\circ$}}
\newcommand{\picplace}[1]{\frame{\centerline{(will be inserted later)}}}
\newcommand{\acknowledgements}{{\it Acknowledgements.}}
\renewenvironment{thebibliography}[1]{{\section*{References}}
\parindent0cm}{\par\vskip1.5em}
\renewcommand{\bibitem}[2]{\par}
\newcommand{\bildh}[3]{{{}}\label{#1}}
\newcommand{\bildw}[3]{{{}}\label{#1}}

\newcommand{\D}{\displaystyle}

\begin{document}

\thesaurus{03(08.09.2 GRS 1915+105; 08.09.2 GRO J1655-50; 11.01.2; 11.10.1;
11.14.1; 02.01.2)}
\title{Galactic jet sources and the AGN connection}
\author{Heino Falcke and Peter L. Biermann}
\offprints{HFALCKE@mpifr-bonn.mpg.de}
\institute{
Max-Planck Institut f\"ur Radioastronomie, Auf dem H\"ugel 69, D-53121
Bonn, Germany}
\date{Submitted to A\&A.}
\markboth{Falcke \& Biermann: Galactic jetsources}
{Falcke \& Biermann:  Galactic jet sources}
\maketitle
\begin{abstract}
In order to further test our hypothesis that jets and disk around
compact accreting objects are symbiotic features we investigate the
newly discovered superluminal galactic radio jets GRS 1915+105 and GRO
J1655-40 and the two famous galactic radio jets 1E1740-2942 and SS 433
within the framework of our couple jet/disk model developed initially
for active galactic nuclei (AGN) and the galactic center source Sgr
A*. By comparing the ``disk'' and radio core luminosity of those
galactic jet sources with our model prediction we can show that they
can easily be understood as AGN-like jets where the accretion power
onto a central compact object is scaled down by several orders of
magnitude.  The total power of the jets must be comparable to the disk
luminosity - at least for the superluminal sources.

To broaden our view we also shortly discuss the situation in other
galactic flat spectrum radio source associated with compact objects --
the X-ray binaries Cyg X-1, Cyg X-2, Cyg X-3 and Sco X-1 -- where a
jet origin has been proposed earlier on theoretical grounds. In an
disk/radio luminosity their radio cores also fall within our model
prediction for scaled down radio loud and radio weak AGN-jets. Taking
all sources together and comparing their $L_{\rm disk}$/radio ratio we
find an indication for a similar radio loud/radio weak dichotomy as
found earlier for quasar radio cores, however, a larger number of
galactic jet sources is needed to confirm this trend.

\keywords{stars: GRS 1915+105 -- stars: GRO J1655-40 -- galaxies: active --
galaxies: jets -- galaxies: nuclei -- accretion}
\end{abstract}
\section{Introduction}
Bipolar outflows with velocities close to the speed of light are found
in the nuclei of many active galaxies (AGN)
and are believed to be powered by a central engine consisting of a
super massive black hole and an accretion disk. In previous papers
we investigated the energy balance of these systems (Falcke, Mannheim,
Biermann 1993, (FMB93); Falcke
\& Biermann 1994 (Paper I); Falcke, Malkan, Biermann 1994, (Paper II);
Falcke, Gopal-Krishna, Biermann 1994) and started with the basic
hypothesis that jets and disks around compact accreting objects are
symbiotic features. We found that a closely coupled jet/disk system
can explain such a variety of phenomena as the UV/radio correlation of
radio loud and radio weak quasars and the unusual radio properties of
the Galactic Center source Sgr A*. Of course it is in principle not
possible to prove the symbiotic nature of jets and disks (i.e. that
they are always both present) but our aim is to check how far we can
get with such an assumption and test it for as many different object
classes as possible.

Recently, it was shown that the relativistic jet phenomenon can be
found inside the Galaxy as well, again in systems which are suspected
to harbor ultra-compact central objects -- most likely stellar mass
black holes (or neutron stars) -- surrounded by accretion disks
(e.g. GRS 1915+105 \& GRO J1655-40). This was not unexpected within
the framework of our approach and hence, we now want to apply our
analysis of jet/disk systems to these objects as well and test if the
previous found $L_{\rm disk}$/radio correlation for AGN extends down
to stellar mass black holes.

The similarity between these galactic jet sources and AGN is often
stressed, but does this also have a physical foundation other than a
pure morphological similarity? Are both engines powered by accretion
onto a compact object producing radio jets?  If AGN and and galactic
jet sources were indeed to be of similar origin, they still should be
understandable within the same simple physical description despite the
scales being vastly different. Such a very simple description was
outlined in Paper I and successfully tested in Paper II for AGN and in
FMB93 for Sgr A* -- already very different systems. There we were able
to account for the observed properties by assuming that only a few
important parameters -- like the mass accretion rate and the black
hole mass -- change, while most others, like the ratio of jet power to
disk luminosity or the fraction of relativistic electrons, remain
largely scale invariant. We also found that indeed mass and energy
conservation are serious constraints for coupled jet-disk systems and
therefore imply interesting conclusions, e.g. about the energization
of electrons. If we were to find similar results for the galactic jet
sources then indeed one had more reason to believe in a physical
relationship between galactic jet sources and AGN. Hence, in this
paper we are not interested in a detailed modelling of individual
sources but want to discuss our scenario introduced in Paper I on the
larger scale.

The plan of the paper is as follows: we first will introduce the
sources we are using (Sec. 2), then discuss the expected values of
$L_{\rm disk}$ (Sec. 3.1) and $L_{\rm jet}$ (Sec. 3.2) for galactic
jet sources using our model description (see Sec. 3.3 for a short
discussion of the parameters) and finally construct a $L_{\rm
disk}/L_{\rm jet}$ diagram for galactic and extragalactic jet sources
followed by a discussion of the results.

\section{The sources}

The sample of galactic jet sources we investigate is not at all
properly selected as we took just the prominent sources known to us
which have received substantial attention in recent years. The sources are
listed in the following.

\paragraph{GRS1915+105} -- the first
galactic jet source found to have superluminal motion (Mirabel \& Rodriguez
1994, hereafter MR94).  This source was studied with the VLA because
of its frequent X- and Gamma-ray bursts. After one of these outbursts
Mirabel \& Rodriguez detected a radio blob obviously being ejected
from the central source. The derived expansion speed was 1.2 $c$. This
phenomenon is well known from extragalactic flat spectrum radio
sources (blazars and core dominated quasars) and usually interpreted
as a relativistic effect caused by a radio jet with bulk velocity
close to the speed of light pointing towards the observer. Thanks to a
sufficent data base collected for the March 19, 1994 outburst MR94
were able to determine the physical properties of the expelled radio
blobs, i.e. a true velocity of $v_{\rm jet}=0.92c$ and inclination of
the jet axis $i=70\circ$ at a distance of $D=12.5$ kpc.

\paragraph{GROJ1655-40} -- an X-ray nova detected with BATSE
in Scorpio which was followed by the expulsion of radio ejecta
apparently at superluminal speeds (Tingay et al. 1995), making this
source the second confirmed superluminal jet in the Galaxy with an
intrinsic speed of the order $\sim0.8$c.

\paragraph{1E1740-2942} -- The first galactic jet source found to
resemble an extragalactic jet was the `great annihilator'
1E1740.7-2942 -- a putative black hole showing a miniature FRII type
jet structure located inside a giant molecular cloud (Mirabel et al.
1992). There is a weak central core and two lobes 1 pc away.  This is
one of the few sources where a broad electron/positron ($e^+e^-$)
annihilation line was claimed to have been observed by the experiment
SIGMA on GRANAT (Bouchet et al. 1991). The canonical model is a
compact object, perhaps accreting from the molecular cloud in an
accretion disk (of mass $M_{\bullet}\sim10M_{\sun}$) which produces a
radio jet flowing along the rotation axis of the system.

\paragraph{SS433} -- is a famous binary system with highly red- and
blue-shifted emission lines due to the ejection of dense gas linked to
a mildly relativistic precessing radio jet moving with a speed of
$v_{\rm jet}=0.26c$ (see Margon 1984). It is often seen as the
archetype for any galactic jet source although it is quite unique in
various respects (i.e. precession of jet axis, weak x-ray flux). It
may well harbor a thick supercritical accretion disk rather than the
usually inferred thin disk. One of the stars is supposed to be either
a neutron star or a black hole.

\paragraph{X-ray binaries} --
There is a whole class of objects, the x-ray binaries, which are
assumed to consist of compact objects (neutron stars or black holes)
and show (variable, flat spectrum) radio emission. Hjellming and
Johnston (1988) proposed that sources like Cyg X-1, Cyg X-2, Cyg X-3,
and Sco X-1 may also have conical jets somewhat similar to
SS433. Therefore, we will also include these sources to compare their
radio/x-ray properties to those of the confirmed jet sources.

\paragraph{Quasars, Sgr A*, M31* } --
We will compare the newly added, low-mass sources with the previously
discussed PG-quasar sample (Paper II) and the nearby supermassive
black hole candidates in the Galaxy (Falcke et al. 1993a\&b, Falcke \&
Biermann 1994, Falcke 1994a\&b) and M31 (M31*) (see Falcke \& Heinrich
1994). The latter do not have confirmed radio jets but compact radio
cores which may be explained as such. The quasar sample includes
typical radio loud FR\,II-type radio quasars and radio weak
quasars. In Paper II we have argued that their radio emission can as
well be attributed to a central radio jet.

\section{$L_{\rm disk}$/radio correlation}

\subsection{Disk luminosity}
The theory of viscous accretion disks is canonically described by the
Shakura \& Sunyaev (1973) $\alpha$-disk model and its relativistic
extension by Novikov \& Thorne (1974). The characteristic effective
temperature of accretion disks $T_{\rm eff}$ is fairly independent of
the details of the viscosity mechanism and merely depends on the mass
of the black hole $M_\bullet$ and the accretion rate $\dot M_{\rm
disk}$, the total disk luminosity $L_{\rm disk}$ depends on $\dot
M_{\rm disk}$ only and we have (see e.g. Falcke et al. 1993)

\begin{equation}
\relax\label{numax} \nu_{\rm max} = 0.7\cdot 10^{18}\> {{\dot {m}_{-8
}^{1/4}} \over {m_\bullet^{1/2}}}{\left( {{ {r^{-3}}}{ {{\cal Q}
{{\cal B}^{-1}{\cal C}^{-1/2}}}}}\right)^{1/4}\over 0.1 }{\rm Hz},
\end{equation}
\begin{equation}\label{teff}
T_{\rm eff}=1.2\cdot10^7 {\rm K} \left({\nu_{\rm max}\over10^{18} {\rm
Hz}}\right)
\end{equation}
and
\begin{equation}
L_{\rm disk} = 3\cdot10^{37} \dot m_{-8}\left({\eta\over5\%}\right)\, {\rm
erg/sec}.
\end{equation}

Here $m=M_\bullet/M_{\sun}$ and $\dot{m}_{-8}=
\dot{M}_{\rm disk}/10^{-8}(M_{\sun}/{\rm yr})$ are the dimensionless mass and
accretion rate of the black~hole. \mbox{$r=R/R_{\rm g}$} is the
dimensionless radius in units of the gravitational radius $R_g=G
M/c^2=1.48\cdot 10^5 M/M_{\sun} {\rm cm}$ which is half the
Schwarzschild radius. The relativistic correction factors ${\cal B,C}$
and ${\cal Q}$ are functions of $r$ and are given explicitly in Page
\& Thorne (1974). The efficiency $\eta$ for black holes varies between
$5\%-30\%$ depending on their angular momentum.

With $T_{\rm eff}$ and $L_{\rm disk}$ one can simply construct a
Hertzsprung-Russell diagram for black holes and their accretion
disks (Falcke et al 1993, Falcke
\& Biermann 1994), where each combination of mass and accretion rate
has its well defined place in the $T_{\rm eff}/L_{\rm disk}$
plane. Quasars with supermassive black holes and high accretion rates
for example can be found around $L\sim10^{44}-10^{48}$ erg/sec and
$T_{\rm eff}\sim10^{4-5}$ K while solar mass black holes have $L_{\rm
disk}\sim10^{36-38}$ erg/sec and $T_{\rm eff}\sim 10^{7}$ K. In order
to discuss a possible $L_{\rm disk}$/radio correlation over such a
broad range of parameters it is not viable to simply compare one small
frequency range with another: this may work for a homogenous class of
objects like quasars, where an UV/radio correlation can be found, but
going from quasars to stellar mass black holes one obviously has to
shift from the UV to the x-ray regime (Eq. \ref{numax}), where the
bulk of the accretion disk luminosity is radiated.

\subsection{Jet-disk coupling}
In a series of paper (FMB93; Falcke et al. 1993a); Paper I; Paper II;
Falcke et al. 1995b; Falcke 1994a\&b) we investigated the correlation
between accretion disk luminosity ($L_{\rm disk}$) and radio
luminosity in quasars as well as in the Galactic Center source Sgr
A*. We found a tight correlation between $L_{\rm disk}$ and radio
emission in radio loud and radio weak quasars (Paper II) and explained
this as the consequence of a closely coupled jet/disk system in the
nuclei of these galaxies. By adding mass and energy conservation in a
jet/disk system to the classical Blandford \& K\"onigl (1977) emission
model for radio cores (Paper I) we showed that any jet/disk coupling
imposes serious constraints on the possible source parameters
(Paper II). The most important finding was that the total jet power
$Q_{\rm jet}$ is a substantial fraction of the disk luminosity $L_{\rm
disk}$ as was already suggested in a seminal paper by Rawlings \&
Saunders (1991).  This suggests that jet formation is directly coupled
to the dissipation mechanism in the disk and occurs very close to the
black hole (Paper I).

If we express the parameters of the jet radio emission in a scale
invariant form and assume that the jet emission is caused by the
overlap of synchrotron self-absorbed components in a conically shaped
jet, fed by the accretion disk, and transporting a tangled magnetic
field gradually declining as $B\propto r^{-1}$ one obtains a flat total
radio spectrum (Paper I, Eq. 57) from one jet cone


\begin{eqnarray}\label{lnu}
L_{\nu_{\rm s, obs}}^*&=& \D{ 2.5\cdot 10^{21}}\,{{\rm erg}\over
{\rm s\, Hz}}\;\left({q_{\rm j/l} L_{38} }\right)^{17/12}\\\nonumber&&\cdot
{{\cal D}^{13/6}\sin i_{\rm obs}^{1/6} \left({\gamma_{\rm e} x_{\rm
e}}\right)^{5/6}
\over \gamma_{\rm j}^{11/6}\beta_{\rm j}^{5/12}u_3^{7/12}}
\end{eqnarray}

Here, it is assumed that magnetic field, turbulent motion and
relativistic particles ($u=3u_3$) are in equipartition
 and form a relativistic plasma with maximal sound speed. The
latter requires also equipartition between kinetic and internal
energy. This {\em maximal} jet (see Paper I) is the radiatively most
efficient type of jet one can get for a given total jet power $Q_{\rm
jet}$, and as we showed in Paper II less efficient jet models fail to
explain the radio emission of radio loud quasars.  We now only assume
that the total jet power scales with the disk luminosity ($Q_{\rm
jet}\la L_{\rm disk}$) and thus obtain Eq. \ref{lnu}.

The parameters introduced are the ratio of jet power to disk luminosity
$q_{\rm j/l}=0.5Q_{\rm jet}/L_{\rm disk}$, the disk luminosity
$L_{38}=L_{\rm disk}/10^{38}$ erg/sec, the ratio between the number
density of relativistic electrons to the number density of protons $
x_{\rm e}$ (without pair creation $x_{\rm e}\le1$), $\gamma_{\rm e}$
is the minimum Lorentz factor of the electrons, $i$ the angle between
jet-axis and the line of sight, $\cal D$ is the Doppler factor of the
jet, $\beta_{\rm j}c$ and $\gamma_{\rm j}$ are bulk velocity and
Lorentz factor of the jet.

The reason for the non-linear dependence of $F_{\nu}$ on $L_{\rm
disk}$ is simply due to the fact that the synchrotron emissivity in
the equipartition case depends non-linearly on the magnetic field
energy density; in addition there is a minor effect induced by
different spectral shapes for different electron energy distributions
which we assume to be a powerlaw with exponent 2.

For an edge-on system of two jet cones in the Galaxy (D=8.5 kpc) we
get (ignoring Doppler boosting and inclination effects, setting
$q_{\rm j/l}=0.15$ as in Paper II)

\begin{eqnarray}\label{flux}
F_{\nu}&=&4\,\mbox{mJy}\cdot L_{38}^{1.4}\left(\gamma_{\rm e}x_{\rm
e}\right)^{.83}\beta_{\rm j}^{0.42}\gamma_{\rm j}^{1.8},
\end{eqnarray}
the physical scale corresponding to this flux and cm wavelengths is

\begin{eqnarray}
z_{\rm jet}(\nu)&\simeq&3\cdot10^{13} \,{\rm cm}\; \left({\nu\over1\,{\rm
GHz}}\right) \left(\gamma_{\rm e}x_{\rm e}\beta_{\rm j}/\gamma_{\rm
j}L_{38}^2\right)^{1/3}
\end{eqnarray}
which is of the right order for radio weak sources. One has to note
that $z_{\rm jet}(\nu)$ corresponds to a distance where the jet
becomes optically thick at the specified frequency. This has to be
taken into account when one discusses radio outbursts and ejection of
blobs. Although an evolving single blob is not really properly
described by our emission model, the best comparison between the radio
flux at one frequency and the model will be possible at that point in
time where the blob spectrum turns over and becomes optically thick at
this frequency. The blob evolution seen by VLBI and VLA usually
depicts later stages of the evolution.

When comparing those highly time dependent events with our stationary
model one also has to consider that there may be different time scales
involved for the jet and the disk (battery effect), hence, the
relation Eq. (\ref{lnu}) may be violated in strong outbursts. In the
comparison of outburst stages with the would-be-stationary calculation
large time scale differences should show up as strong deviations from
the model.

\subsection{Parameters}
We now shortly discuss the various parameters entering the model:

\subsubsection{Total jet power}
For quasars we found (Paper II) that the ratio between jet power and
disk luminosity was around $2 q_{\rm j/l}=0.3$. Substantially lower jet
powers were not able to explain radio loud jet cores -- even with the
most efficient models. As we assume that this parameter reflects an
universal jet/disk coupling mechanism, we will keep that value fixed
in the further discussion.

\subsubsection{Jet velocity}
In Paper II we discussed the possibility of a scaling of the proper
jet velocity with disk luminosity but did not find a strong effect. However,
we found that a weak power-law dependence with exponent $\xi=0.15$ is
quite consistent with the data (Paper II, Eq. 11). In contrast to
quasars we know the jet velocities in the galactic superluminal
sources quite well which are in the range (0.7-0.9)c and we can fix
the velocities at the measured values.  To faciliate a smooth
transition from AGN to galactic jets we therefore slightly modify the
scaling law to

\begin{equation}
\gamma_{\rm j}\beta_{\rm j}=1.3 \left(1+{L_{\rm disk}\over10^{43} {\rm
erg/sec}}\right)^{0.15}.
\end{equation}

The previously used pure powerlaw would have predicted $\beta_{\rm
j}\sim0.4$ for systems with $L_{\rm disk}\sim10^{38}$ which is still
in the right ball-park but slightly too low to produce the
relativistic effects in GRS 1915+105 and GRO J1655-40. At present we
do not attribute any physical significance to this scaling law, use it
mainly for practical reasons and will not adjust the correct value for
each individual source as the changes due to a different jet speed in
the regime $\beta\la0.8$ are smaller than the overall uncertainties
of the model.

\subsubsection{Electron content: radio loud -- radio quiet}
Choosing a relativistic electron content of the jet with $\gamma_{\rm
e}x_{\rm e}\sim1$ would correspond to a situation where the jet is
already very energetic, however, the radiative efficiency is limited
by the number of available electrons in the flow which have all been
accelerated into a power law distribution. Neither magnetic field nor
the number of electrons can be further increased without violating
energy and mass-conservation in the jet/disk system to produce more
radio emission even though the electrons are not yet in
equipartition. One can only alter the distribution of the electrons by
raising their number by additional pair-creation ($x_{\rm e}\la100$,
e.g. mixed or pure electron/positron jet) or injecting the electrons
already at a high energy where they become energetically dominant
($\gamma_{\rm e}\la100$, see Paper I). The latter is what must happen
somehow in radio loud quasars (Paper II) and applied to
Eq. (\ref{flux}) means that one may expect fluxes of $\sim100$ mJy up
to a few Jy for the Eddington limit of stellar mass black holes.

In Paper II we tentatively identified a maximal jet with $\gamma_{\rm
e}=1$ \& $x_{\rm e}=1$ (all electrons accelerated from thermal pool
but only protons are in equipartition with the magnetic field) as the
radio quiet state and $\gamma_{\rm e}=100$ (electrons and protons in
equipartition with magnetic field, electrons injected at high
energies) as the radio loud state -- both are natural states within
any jet model. The radio loud model is the radiatively most efficient
situation and for a given $Q_{\rm jet}/L_{\rm disk}$ also is the upper
limit to the possible radio emission - one would strongly suspect that
$Q_{\rm jet}/L_{\rm disk}\la1$ in a normal system.

One way to obtain the injection of synchrotron radiating particles at
$\gamma_{\rm e}\sim100$ and make jets radio loud can be the pion-decay
in hadronic cascades which in turn could be initiated by $pp$ collisions
of relativistic protons in the jet with thermal protons in the shear
layer between jet and surrounding medium (see Paper I\&II, Falcke
1995).

\section{A universal $L_{\rm disk}$/radio correlation?}
\begin{figure*}
\centerline{\bildh{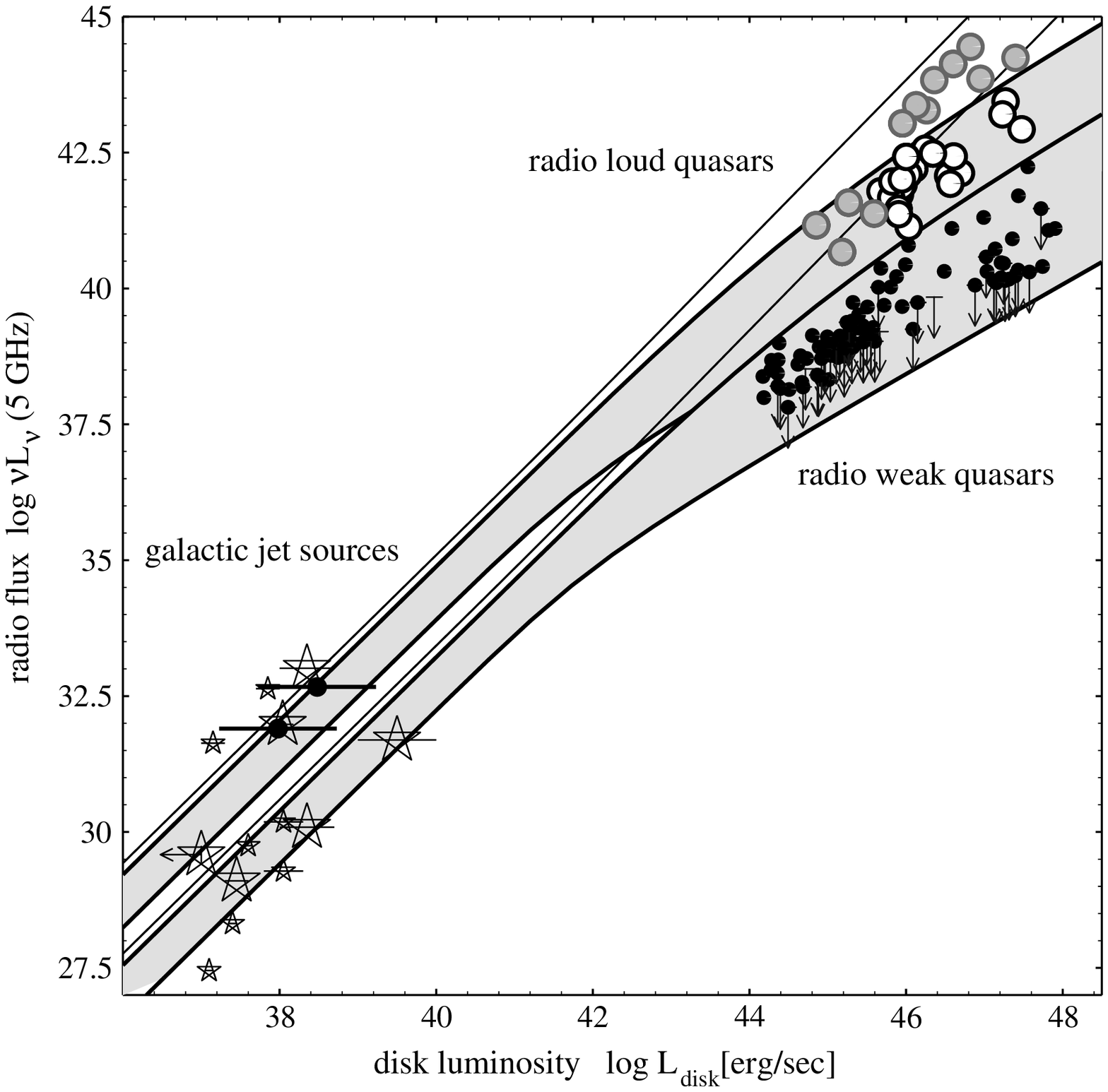}{12cm}{bbllx=1.3cm,bblly=4.8cm,bburx=19.2cm,bbury=21.9cm}}
\caption[]{\label{correlation}
Monochromatic luminosity of the compact radio core versus the disk
luminosity for AGN (above $10^{43}$ erg/sec, see Paper II for details)
and galactic jet sources (stars).  The two dots among the galactic jet
sources are Sgr A* and M 31*, the x-ray binaries are represented by
smaller stars.  The shaded band gives our jet model for inclination
angles of $15\degr$ to $90\degr$, the upper lines indicate emission
seen within the boosting cone (0-15 degree).}
\end{figure*}

\subsection{Application of the model}
We now want to apply our scaling laws (Eq. \ref{lnu}) to the galactic
radio jets and compare them with other known jet sources. We will take
the same model as in Paper II with just the parameters used to
describe the radio/UV correlation in quasars (``maximal'' radio loud
and radio weak jets with $q_{\rm j/l}=0.15$ and $\gamma_{\rm
jet}(L_{\rm disk}=10^{46}{\rm erg/sec})=6$).

The width of the model prediction (Fig. \ref{correlation}) is
determined by the spread of possible inclination angles in conjunction
with relativistic boosting. The shaded band limits the range between
15 and 90 degree, while the thin line represents the 0 degree
inclination case. As this spread is smaller for lower jet velocities
($\beta<0.8$) those models are simply contained within the range of
the presented model.

One should remember that for quasars the lower part of each band is
probably not populated because of obscuration by a torus. The bend in
the bands reflects the turnover from highly to mildly relativistic
jets. Figure \ref{correlation} was already presented in Falcke (1994)
-- before the discovery of the superluminal sources and the
determination of their velocities -- for slightly lower jet
velocities, but despite the bands being somewhat narrower it already
re\-presented the same basic results.

\subsection{Observational data for individual sources}
The next step is to compare $L_{\rm disk}$ and radio core fluxes
for the different object types with the basic jet model.  We first
reproduced Fig. 3a from Paper II, where we compared UV-bump and VLA
radio cores of a quasar sample with the model of Paper I, and extended
the plot range to very low disk luminosities.  Moreover, we  added Sgr A*
and M31* as already discussed in Falcke et al. (1993a), FMB93 and
Falcke \& Heinrich (1994).

We then estimated the disk luminosities for the galactic jet sources
GRS 1915+105, GRO J1655-40, and 1E1740-2942 from their X-ray
observations where we differentiated between high and low states if
required. The accretion disk luminosity of SS433 was already
extensively discussed in the literature. Those numbers are given in
Table 1 and more detailed comments and references for the individual
sources are given in the appendix. In additoin we collected the 5GHz
flat spectrum core fluxes of the jets corresponding to the sources in
their different states as listed in Table 1 and discussed in the
Appendix. For the outburst sources we either took the peak x-ray
luminosities and the flux when the spectrum appeared flat at 5 GHz in
an outburst. We then did the same for the X-ray binaries. When several
datapoints of a variable source were available we choose the
logarithmic mean value of the radio flux (in low and high states
respectively) as we did in Paper II already.

\subsection{Results}

The combined plot of these sources together with the model is shown in
Figure \ref{correlation}. One can see that in fact most sources
directly fall in the predicted range and none really is far off, only
Sco X-1 seems to have a somewhat low radio flux compared to ist x-ray
luminosity. The sources also seem to be divided into two branches
separated by roughly 2-3 orders of magnitude.

{\em Considering the uncertainties we at least can positively state that
each of those sources could in principle be explained by a jet model
within the constraints of mass and energy conservation in a coupled
jet/disk system as outlined in Paper I.}

\begin{figure}
\centerline{\bildw{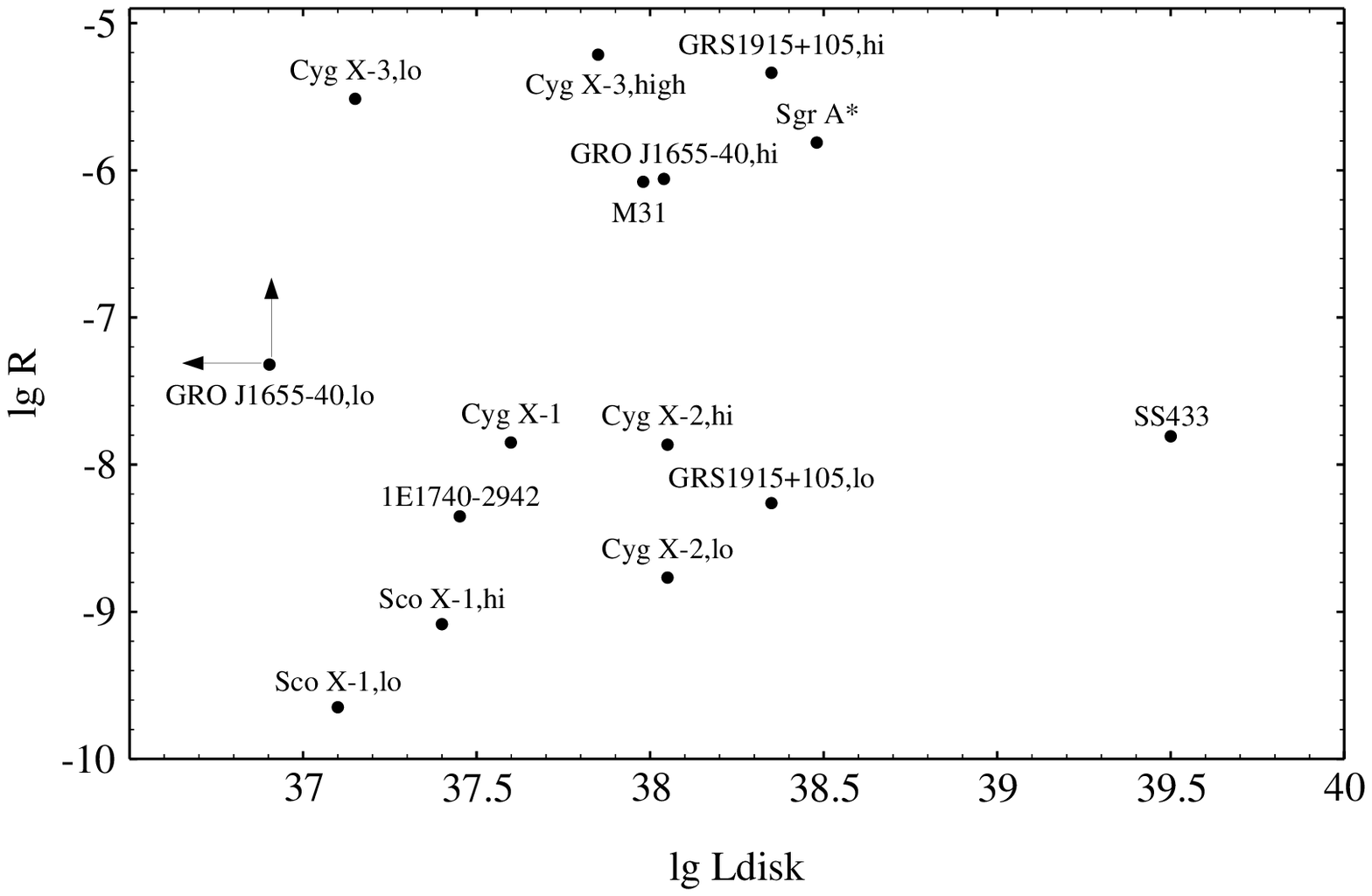}{8.7cm}{bbllx=2.5cm,bblly=8.5cm,bburx=18.9cm,bbury=19.4cm}
}
\caption[]{\label{logR} The logarithm of the ratio $R$ between radio
luminosity at 5 GHz and disk luminosity for the galactic jet sources.}
\end{figure}

To highlight the separation of the two branches, we calculated the
ratio between the disk luminosity and the radio luminosity at 5 GHz
for each source (Table 1), yielding something like the $R$ parameter
used in the description of radio loudness of quasars.  We plotted this
$R$ parameter versus the disk luminosity of the sources (Fig.
\ref{logR}); there indeed appears to be  a separation into
a radio loud branch with $\log R>-6$ and a radio weak branch with
$\log R<-8$.

In Fig. \ref{logR} we also labeled the individual sources which
makes it easier to also locate these sources in Figure
\ref{correlation}.  Most x-ray binaries Sco X-1, Cyg X-1, Cyg X-2 in
high and low states fall in the radio weak regime occupied by
1E1740-2942 and SS433. On the other hand Cyg X-3 falls on the radio
loud branch occupied by the starved supermassive black holes Sgr A*
and M31* and the superluminal jets GRO J1655-40 and GRS 1915+105. The
latter are located at the uppermost end of the model which is
consistent with them being relativistically boosted. Having the
warning in mind that for this source we compare a time dependent
outburst with a stationary model we may have another factor besides
boosting that may lead to an enhancement of the radio
emission. In the low states both jet sources are well within the model
predictions where the low radio state of GRS 1915+105 even falls
completely onto the radio weak branch while the case for GRO J1655-40
remains less clear due to the non-detection in x-rays. Cyg X-3 remains
radio loud even in the low state.

\begin{table*}
\begin{tabular}{|l|c|c|c|c|c|c|}
\hline
Object & state & D [kpc]  & $F_{\rm 5GHz}$ mJy &
$L_{\rm disk}$ [erg/sec] & lg $R$\\
\hline\hline
GRS 1915+105 & low      & 12.5 &  $\sim1$  & $3\cdot10^{38}$&-8.3\\
             & outburst &      &  655  & $3\cdot10^{38}$&-5.3\\
GRO J1655-40 & low      & 4    &  4    & $<8\cdot 10^{36}$&$>$-7.4\\
             & outburst &      &  1000 & $      10^{38}$&-5.3\\
1E1740-2942  & normal   & 8.5  & 4     & $3\cdot10^{37}$&-8.4\\
SS 433       & normal   & 4.85 & 350   & $1-10\cdot10^{39}$&-7.8\\
Cyg X-1      & normal   &2.5   & 15    & $4\cdot10^{37}$&-7.8\\
Cyg X-2      & low      & 8    & 0.5   & $1\cdot10^{38}$&-8.8\\
             & high     &      & 4     & $1\cdot10^{38}$&-7.9\\
Cyg X-3      & low      & 8.5  & 100   & $1\cdot10^{37}$&-5.5\\
             & high     &      &$10^{3-4}$& $1\cdot10^{38}$&-5.3\\
Sco X-1      & low      & .5   & 10    & $1\cdot10^{37}$&-9.6\\
             & high     &      & 1     & $3\cdot10^{37}$&-9.1\\
\hline
\end{tabular}
\caption[]{Parameters of the considered objects. Col.(1): object
name, Col. (2): state of the object (high, low, outburst), Col. (3):
distance assumed, Col. (4):
estimated mean disk luminosity, Col. (5): mean radio flux at 5 GHz,
Col. (6): log of ratio between radio luminosity at 5 GHz and disk
luminosity}
\end{table*}

\section{Discussion}
In Paper I we derived a simple jet model by adding mass and energy
conservation in a jet/disk system to a simple sychrotron emission
model of jets first developed by Blandford \& K\"onigl (1979). This
model is derived from first principles and is fairly universal as it
simply describes the dependence of a synchrotron emitting jet source
on simple plasma parameter. By parametrization of this model in terms
of the power available in an accretion disk thought to power the jet,
we obtained a model which does not have an intrinsic scale other than
the power of the accretion disk. Therefore we suggested that it should
be applicable to all types of radio jets. And indeed it was possible
to use the same kind of model for a low power jet source like Sgr A*
and and high power sources like quasars.

We have now extended this analysis and compared the overall radio
properties of the cores of galactic jet sources and x-ray binaries
with their disk luminosity.  We again find a trend of increasing radio
luminosity with increasing disk power.  There also appears to be a
separation in radio loud and radio weak sources. While the radio loud
sources have an $R$-ratio between radio luminosity at 5 GHz and disk
luminosity of $R\sim10^{-5..-6}$, radio weak sources have
$R\sim10^{-8..-9}$. Given the large offset between radio loud and
radio weak this dichotomy should prevail even if some of the
measurements were in error, but considering the small number of sources
and their ill defined selection criterion we can not exclude that this
is caused by a selection effect. Nevertheless, this is reminiscent of the
situation in quasars and may be an interesting trail to be followed in
future research.

So far GRO J1655-40, Cyg X-3, Sgr A*, and M31* appear to be
radio loud, while Sco X-1, Cyg X-1, Cyg X-2, 1E1740-2942 and SS433
appear radio weak.  GRS 1915+105 is the only source that varied by a
factor 1000 while its x-ray luminosity was roughly constant and
therefore moved from the radio loud branch in the high state to the
radio weak branch in the low state.

Comparing the luminosities with our model prediction we found that,
like quasars, galactic jet sources and our x-ray binaries can as well
be explained within the same simple physical description of an
outwards moving, synchrotron emitting plasma in a radio jet being fed
by an accretion disk. Interestingly, the same intrinsic model
parameters as used for extragalactic jets have to be used for galactic
jet sources as well. These parameters are the ratio between jet power
and disk luminosity which must be very high ($Q_{\rm jet}/L_{\rm
disk}\sim0.3$), the minimum Lorentz factor of the radiating electrons
which has to be $\sim100$ for the radio loud sources (or alternatively
100 fold more particles by pair creation) and the state of
the plasma which has to be near relativistic and near its
equipartition value (between magnetic field and relativistic
particles) - only jet velocities seem to be slightly smaller in low
power jets than in quasars. Thus the radio loud galactic jet sources
are already in the most efficient state as far as their radio
emission is concerned which means that we have no freedom any more in
choosing the plasma parameters.

The similarity in the physical conditions in galactic and
extragalactic jets shows that they very likely have also a physical
link beyond mere morphological similarity -- they seem to have
similar engines differing only in size and power. Thus it is no
coincidence that both types of engines are usually interpreted as
compact accreting objects.  In our formulation the nature of this
compact object is irrelevant and should be valid for black holes,
neutron stars and perhaps even T Tauri stars. However, if like in
stellar wind theory, the terminal jet velocity scales somehow with the
escape velocity at the sonic point -- which usually is close to the
base of the jet/wind -- the terminal velocity could be a function
primarily of the compact object. Rapidly rotating Kerr holes with high
escape speeds would thus require relativistic jets while Neutron stars
or Schwarzschild holes with lower escape speeds would require only
mildly relativistic jets, stars would produce jets with only
subrelativistic speeds. As the escape velocity close to a black hole
at a fixed dimensionless radius $r=R/(GM_\bullet/c^2)$ is the same
irrespective of the central mass $M_\bullet$ we could understand why
the jet velocities in galactic and extragalactic jets are so similar
despite 8-10 powers of ten difference in mass and mass accretion
between both types.

One may speculate -- as we did in Paper I -- that indeed jets are
quite naturally associated with compact accreting objects but come in
two flavors -- radio weak and radio loud -- depending on the
energization of electrons.  We propsed previously that those states
could correspond to the situation where a) the thermal electrons
($\gamma_{\rm e}\sim1$) are smoothly accelerated into a powerlaw
distribution b) electrons (or pairs) are shifted upwards in energy
until the equipartition value is reached yielding a minimum Lorentz
factor for the electrons of $\gamma_{\rm e}\sim100$. We note that one
process that can inject pairs at such a high energy is the pion decay
in hadronic cascades (see Falcke 1995, Falcke et al. 1995, Biermann et
al. 1995 for examples).

This connection between galactic and extragalactic jets is highlighted
in Fig. \ref{correlation} where the same jet emission model smoothly
connects high power and low power sources. Although we still do not
have enough sources to firmly claim a $L_{\rm disk}$/radio correlation
for galactic jet sources the results are already very promising.  The
hunch of a universal $L_{\rm disk}$/radio correlation for compact
accreting objects extending from AGN down to stellar mass black holes
as implied by Paper I\&II and FMB93 is still consistent with all
sources we tested. A future step to firmly establish such a
correlation will be to investigate supermassive black hole candidates
at intermediate powers between stellar mass black holes and AGN in
nearby galactic nuclei and low-power Seyfert galaxies. In this sense
one should understand Fig. 1 as a prediction for these sources.

It is interesting that also the sources where it is not yet rigorously
shown that the radio emission is caused by jets like Sgr A*, M31*, Cyg
X-3 and the other x-ray binaries mix with those sorces where there
obviously are jets (SS433, 1E1740-2942, GRO J1655-40, GRS
1915+105). This shows that the former can also easily be explained by
a simple jet model but does of course not yet prove the existence of
jets there. The observation that Cyg X-2 moves up and down within the
radio weak band indicates that our treatment is indeed too simplified
for a detailed discussion of the sources but on the large scale we still
expect that further putative galactic jet sources will fall within the
bands predicted by our model.

We finally conclude that the comparison of the flat spectrum core flux
of compact accreting objetcs with their putative disk luminosity --
the peak in the spectral energy distribution -- may become a very powerful
tool for the analysis and classification of such systems. What has to
be done next is to extend the $L_{\rm disk}$/radio data set for
compact accreting objects of stellar mass size in high and low states,
and closely monitor the evolution of the superluminal jet sources to
prove or disprove the notion of a universal $L_{\rm disk}$/radio
correlation and the possible radio loud/radio weak dichotomy for
galactic jet sources.


\acknowledgements
HF is supported by a Max-Planck Stipend. We thank M. Ostrowski for his
critical and helpful comments.

\appendix
\section{Appendix}
The data used for the individual sources are discussed in the
following:

\paragraph{GRS 1915+105}
GRS 1915+105 was only recently discovered as an important object and
has only a limited number of observations available. Using the VLA at
3.6 cm MR94 measure a 8 GHz flux of the slightly resolved source of
655 mJy at March 24, 1994 corresponding to 841 mJy at 5 Ghz for the
spectral index 0.49. This was right after the ejection of the radio
blobs. The same authors earlier reported a somewhat weaker outburst
(IAUC 5900) where the spectrum peaked with 590 mJy at 5 GHz while it
normally showed emission of only a few mJy or even less (IAUC 5773,
5830).  Hence $L_{\rm disk}\sim3\cdot10^{38}$ (MR94) and $F_{\mbox{5
GHz}}\sim655 mJy$ seem to be reasonable numbers to describe GRS
1915+105 in an outburst state. In its lowest state GRS 1915+105 showed
an apparently flat (or inverted) spectrum with as flux as low as 0.2
mJy at 3.6 cm on April 27, 1993 (IAUC 5830) while the X-ray flux was
even somewhat brighter than in March 1994 (Brandt et al. 1993, IAUC
5779; Sazonov et al. 1994, IAUC 5959). We adopted a distance of 12.5 kpc.

\paragraph{GRO J1655-40}
We will concentrate on the August 1994 outburst of this source which was
covered by the VLBI observation and showed superluminal motion; we
will adopt a distance to the source of 4 kpc (Tingay et al. 1995,
Bailyn et al. 1995).  The BATSE team (Zhang et al. 1994, IAUC 6046)
reported an 20-200 keV x-ray luminosity of $2.5\cdot10^{37}$ erg/sec
during the outburst, given the steep photon spectrum of $\sim -3$ the
total x-ray luminosity must have been substantially higher and of the
order $\sim10^{38}$ erg/sec as quoted by Harman et al. (1995) and Tingay
et al. (1995).

The radio outburst started after August 12 with a 5 GHz flux of 1.1 Jy
and 2.5 Jy on Aug 18. The 843 MHz flux peaked on Aug 15 at 5.5 Jy
compared to 0.9 Jy on Aug 12 where the source actucally showed a flat
spectrum (Campbell-Wilson \& Hunstead 1994, IAUC 6052, 6055 \& 6062) ;
Hjellming 1994, IAUC 6055 \& 6060). Hence we used in
Fig. \ref{correlation} the geometric mean flux between 1 and 5 Jy in
Fig. \ref{correlation} as an average value for the outburst. Several
outbursts followed thereafter which reached similar if somewhat lower
numbers. We can catch the low state of this source by looking at the
situation prior to an outburst where the lowest flux levels were
reached but a flat radio spectrum is detected. Such a situation
occurred on Nov. 1 where the hard x-ray luminosity had decreased to
less than $\sim2\cdot10^{36}$ erg/sec (hence a total x-ray luminosity
of $\la8\cdot 10^{36}$ erg/sec (Zhang et al. 1994, IAUC 6101) and the
radio flux to 4 mJy at 5GHz with a flat high frequency spectrum
(Hjellming et al. 1994., IAUC 6107).

\paragraph{1E1740.7-2942}
is a strong emitter of hard x-rays and was observed in two
states by the russian spacecraft GRANAT on various occasions (Bouchet
et al 1991, Sunyaev et al 1991). In its `standard state' the
luminosity in the 4-300 keV band where most of the luminosity is
emitted was $L_{\rm x}=3.2\cdot10^{37}$ erg/sec. There is also a hard
state with an additional broad peak around 480 keV (redshifted pair
annihilation?) and a report of low state with only $L_{\rm
x}\sim2\cdot10^{36}$ erg/sec. We only consider the normal state with
luminosity $L_{\rm disk}=2-4\cdot10^{37}$
erg/sec. Chen et al. (1994) also argue for an average disk luminosity
of $3\cdot10^{37}$ erg/sec (D=8.5 kpc).

The VLA observations of Mirabel et al. (1992) showed a beautiful edge
brightened extended jet structure of 1E1740.7-2942 with a central radio
core possibly having a flat spectrum. Compared to the x-ray flux the
source is very weak in radio and the average VLA flux of the core at
5GHz is around 0.3 mJy.

\paragraph{SS 433}
This source does not quite fit into the ``Hertzsprung-Russell diagram
for black holes'' as its primary emission component which is
associated with the accretion disk does not radiate in x-rays as one
would expect for the low central mass of a few solar mass
(Eq.\ref{numax}) but in the UV as do AGN. This, however, is not too
surprising if one remembers that SS 433 is in a close binary system
and quite probably surrounded by a thick, super-critical accretion torus
and therefore may have completely different radiation characteristics
then a `normal', sub-critical thin disks. Still energy- and
mass-conservation must be obeyed and we can include it into our
investigation. The distance we are using is 4.85 kpc (Vermeulen et
al. 1993a).

The bolometric disk luminosity of SS 433 is usually estimated to be in
the range $L_{\rm disk,S433}\sim10^{39}-10^{40}$ erg/sec (Wagner 1986,
Anthokina \& Cherepashchuk 1987). The discussion of the corresponding
radio data is somewhat more complicated as it requires to disentangle
the various contributions of this highly variable source. Fortunately
there was a large campaign dedicated to SS 433 in May/June 1987
including VLBI and multi frequency radio-photometric observations
(Vermeulen et al. 1993a\&b). The VLBI data show a central core with 5
GHz fluxes in the range 22-190 mJy and total fluxes between 102 and
411 mJy, from the monitoring data one finds that flat spectrum flares
can reach up to 800 mJy. One might label all three contributions as
coming from the core and to be consistent with all other datasets and
our model we will stick with the (geometric) mean VLA core fluxes
(Hjellming \& Johnston 1981) as used in Paper II yielding $\sim 0.35$
Jy for SS 433, quite consistent with the total flux of the VLBI
structure.

\paragraph{X-ray binaries}
Cyg X-1 shows a relatively constant mean radio flux with a flat
spectrum around 15 mJy (Hjellming et al 1975) and an average x-ray
luminosity of $4.2\cdot10^{37}$ erg/sec (Liang \& Nolan 1984), the
variations in both bands are usually less than a factor 2.

Cyg X-2 has a low state and a flaring state roughly corresponding to 5
GHz fluxes between 0.5 mJy and 4 mJy with peaks up to 12 mJy
(Hjellming et al. 1990). The x-ray luminosity is on average $10^{38}$
erg/sec where the variation may yield a factor 2 (Hasinger et al.
1990).

Cyg X-3 does show relativistic expansion with $v\ga0.35c$ (Geldzahler
et al. 1983, Spencer et al. 1986), a high x-ray variability with a low
state around $10^{37}$ erg/sec and a high state with up to $10^{38}$
erg/sec (Watanabe et al. 1994). There seems to be a correlation between
the x-ray flares and strong radio outbursts. Which may reach fluxes up
to 10 Jy and average around 1 Jy, while the quiescence level is around
100 mJy (using a flat spectral index).

Sco X-1 again shows an x-ray variability of a factor 2, with an
average x-ray luminosity of $\sim10^{37}$ erg/sec (e.g. Hasinger 1987)
at and adopted distance of 500 pc. The VLA radio core flux is around 1
mJy in the low state while the state is between 10-20 mJy (Geldzahler
\& Formalont 1986). There is a correlation between  x-ray and radio
flaring (Hjellming et al. 1990).

\end{document}